\documentstyle[11pt]{article}
\def\ft#1#2{\textstyle{{\scriptstyle #1}\over {\scriptstyle #2}}}
\def\st#1{{\scriptstyle #1}}

\def\ww3{{$W_3$}}

\def\del{\partial}
\def\a{\alpha}

\begin{document}
\topmargin 0pt
\oddsidemargin 5mm
\begin{titlepage}
\begin{flushright}
CTP TAMU-10/95\\
hep-th/9503159\\
\end{flushright}
\vspace{1.5truecm}
\begin{center}
{\bf {\Large Higher-spin Realisations of the Bosonic String}}
\footnote{Supported in part by the
U.S. Department of Energy, under grant DE-FG05-91-ER40633}
\vspace{1.5truecm}

{\large H. L\"u,  C.N. Pope and K.W. Xu}
\vspace{1.1truecm}

{\small Center for Theoretical Physics, Texas A\&M University,
                College Station, TX 77843-4242}

\end{center}

\vspace{1.0truecm}

\begin{abstract}
\vspace{1.0truecm}

     It has been shown that certain $W$ algebras can be linearised by the
inclusion of a spin--1 current.   This provides a way of obtaining new
realisations of the $W$ algebras.  Recently such new realisations of $W_3$
were used in order to embed the bosonic string in the critical and
non-critical $W_3$ strings.   In this paper, we consider similar embeddings
in $W_{2,4}$ and $W_{2,6}$ strings.   The linearisation of $W_{2,4}$ is
already known, and can be achieved for all values of central charge. We use
this to embed the bosonic string in critical and non-critical $W_{2,4}$
strings.  We then derive the linearisation of $W_{2,6}$ using a spin--1
current, which turns out to be possible only at central charge $c=390$.  We
use this to embed the bosonic string in a non-critical $W_{2,6}$ string.

\end{abstract}
\end{titlepage}
\newpage
\pagestyle{plain}

       With the discovery of the property of the $W_3$ algebra that it can
be linearised by the inclusion of a spin--1 current \cite{ks1}, new
realisations were constructed for the purpose of building the corresponding
$W_3$ strings \cite{lpsx,bn}.  An unusual feature of these realisations is
that the spin--3 current contains a term linear in a ghost-like field.  The
realisations also close when this term is omitted, under which circumstance
the corresponding string theory is equivalent to the one that is based on
the Romans' free-scalar realisation \cite{romans}.  However, when this term
is included, the corresponding BRST operator is equivalent to that of the
bosonic string, which can be shown by making a local canonical field
redefinition \cite{lpsx}.  Thus the new realisations provide embeddings of
the bosonic string in the $W_3$ string.

    It is interesting to generalise the above consideration to the embedding
of the bosonic string in $W_{2,s}$ strings, where $W_{2,s}$ denotes the
conformal algebra generated by a spin--$s$ current together with the
energy-momentum tensor.   The $W_{2,s}$ strings based on free-scalar
realisations were extensively discussed in Ref.\ \cite{w2s}, where it was
shown that when $s\ge 3$ the cohomologies describe Virasoro strings coupled
to certain minimal models.  The $W_{2,s}$ algebras exist at the classical
level for all positive integer values of $s$.  However, at the quantum
level, for generic values of $s$, a $W_{2,s}$ algebra exists only for a
finite set of special values of central charge \cite{kw,nahm}, which in
particular does not include the critical value.   The exceptions are the
$W_{2,s}$ algebras for $s=1,2,3,4$ and 6, for which the central charge can
be arbitrary.   Although the $W_{2,s}$ algebra does not close at the
critical central charge for generic values of $s$, it is nevertheless
possible to build $W_{2,s}$ strings with free-scalar realisations.  It was
shown in Ref.\ \cite{w2s2} that one can first use the free-scalar realisation
to write down the classical BRST operator, and then quantise the theory by
renormalising the transformation rules and adding necessary quantum
counter-terms.

      The new realisations that were constructed in Ref.\ \cite{lpsx}, which
provide embeddings of the bosonic string in $W_3=W_{2,3}$ strings,
do not generate the $W_3$ algebra at the classical level.  The $W_3$ symmetry
arises only as a consequence of quantisation.   Thus it seems that if we are
to use such new realisations for values of $s$ other than 3, we must
restrict our attention to the cases $s=1, 2, 4$ and 6, for which the quantum
algebras exist.   The embeddings of the bosonic string in the $W_{2,s}$
string for $s=1, 2$ and 3 were discussed in Ref.\ \cite{lpsx}.   In this
paper, we shall focus our attention on the remaining cases $s=4$ and 6.

      It is instructive to begin by studying the form of the linearisation
of the $W_3$ and $W_{2,4}$ algebras, for which the results were obtained in
Refs.\ \cite{ks1,ks2}.  The associated linearised $W_{1,2,3}$ and
$W_{1,2,4}$ algebras take the form:
\begin{eqnarray}
T_0(z)\, T_0(0)&\sim& {c\over 2 z^4} + {2 T_0\over z^2} + {\del T_0\over z}
\ ,\qquad T_0(z)\, W_0(0) \sim {s W_0\over z^2} + {\del W_0\over z}\ ,
\nonumber\\
T_0(z)\, J_0(0)&\sim& {c_1\over z^3} + {J_0 \over z^2} + {\del J_0\over z}
\ , \qquad J_0(z)\, J_0(w)\sim -{1\over z^2}\ ,\label{genform}\\
J_0(z)\, W_0(w)&\sim& {h W_0\over z}\ , \qquad W_0(z)\, W_0(w)\sim 0\ ,
\nonumber
\end{eqnarray}
where $s=3$ and 4 respectively. The coefficients $c, c_1$ and $h$ are given
by
\begin{eqnarray}
c=50 + 24t^2 + {24\over t^2}\ ,&& c_1=-\sqrt{6}(t+{1 \over t})\ ,
\qquad h=\sqrt{\ft32} t\ ,\qquad
(s=3)\nonumber\\
c=86 + 30 t^2 + {60\over t^2}\ ,&& c_1=-3t-{4\over t}\ , \qquad h=t\ .
\qquad\qquad (s=4)\label{coefsalg}
\end{eqnarray}
The currents of the $W_3$ and $W_{2,4}$ algebras are then given by
\begin{equation}
T=T_0\ ,\qquad W=W_0 + W_{\rm R}\ ,\label{realab}
\end{equation}
where $W_{\rm R}$ is the Romans type realisation constructed from $T_0$ and
$J_0$.  For the cases where $s=3, 4$ and 6, $W_{\rm R}$ takes the form
\cite{w2s2}
\begin{equation}
W_{\rm R} = \sum_{n=0}^{[s/2]} g_n(s) J_0^{s-2n}\, T_0^n + {\rm quantum}\,\,
{\rm corrections}\ ,
\end{equation}
where the $g_n$'s are given by
\begin{equation}
g_n={(-2)^{s/2} (s-n-1)!\over 2^n n! (s-2n)!}\ .
\end{equation}
The coefficients $c$ and $c_1$ can be determined from the background charges
of the free-scalar realisation.  In terms of the two scalar realisation, the
energy-momentum tensor can be expressed as
\begin{equation}
T_0=-\ft12 (\del\vec \phi)^2 - (t\vec\rho + {1\over t} \vec\rho^\vee)\cdot
\del^2 \vec \phi\ ,\label{twoscalart}
\end{equation}
where $\vec\phi=(\phi_1, \phi_2)$.  The vectors $\vec\rho$ and
$\vec\rho^\vee$ are the Weyl vector and co-Weyl vector for the Lie algebras
$A_2$ and $B_2$, associated with $W_3$ and $W_{2,4}$ respectively.  For the
$A_2$ algebra, we have $\vec\rho=\vec\rho^\vee = (\sqrt{\ft32},
\sqrt{\ft12})$; for $B_2$, $\vec\rho= (\ft32,\ft12)$ and $\vec\rho^\vee
=(2,1)$.  In each case, the scalar $\phi_2$ occurs in the higher-spin
current only {\it via} the energy-momentum tensor $T_0$, and hence its
contribution can be replaced by an arbitrary effective energy-momentum
tensor.  The scalar $\phi_1$ thus plays a distinguished r\^ole.  It has a
background charge $\a = -\sqrt{\ft32} (t+t^{-1})$ for $W_3$ and $\a=-\ft32 t
- 2t^{-1}$ for $W_{2,4}$.   Thus $c_1=2\a$ in each case.  The central charge
$c$ follows immediately from Eqn.\ (\ref{twoscalart}).   That $c_1=2\a$
is not coincidental.  In fact one can realise the algebra given in Eqn.\
(\ref{genform}) by the two scalar realisation, with $J_0=\del\phi_1$,
$W_0=0$ and $T_0$ given by Eqn. (\ref{twoscalart}).   The third-order pole
in the OPE $T_0(z)J_0(0)$ is then precisely $2\a$.

     We now turn to the case of $W_{2,6}$.  With the coefficients $c, c_1$
and $h$ undetermined, the form given by Eqn.\ (\ref{genform}) with $s=6$ is
the most general possible for the linearised $W_{1,2,6}$ algebra. {\it A
priori}, one might have expected, without loss of generality, that there
could be linear terms involving the currents $J_0, T_0$ and their
derivatives in the OPE $J_0(z)W_0(0)$.  However, we have verified that these
terms are excluded by the requirement of closure of the $W_{1,2,6}$ algebra.
 To determine the coefficients $c, c_1$ and $h$, we use the realisation
(\ref{realab}) to construct the $W_{2,6}$ quantum algebra.  The most general
form for $W_{\rm R}$ in this case has 29 terms.  The requirement that $W$ in
Eqn.\ (\ref{realab}) be primary determines all but three of the associated
coefficients.  The remaining coefficients can be determined by studying the
OPE $(W_0(z) W_{\rm R}(0) + W_{\rm R}(z) W_0(0))$, in which all terms
involving $J_0$ have to be zero.   This determines all the rest of the
coefficients, including $c, c_1$ and $h$.  Unlike the previous cases of
$W_{2,3}$ and $W_{2,4}$, where the coefficients $c, c_1$ and $h$ are
expressed in terms of the free parameter $t$ in Eqn.\ (\ref{coefsalg}), here
these coefficients are uniquely determined, modulo a trivial reflection
symmetry $J_0 \longrightarrow -J_0$, namely $c=390, c_1 = 11$ and $h=-1$.
One might have expected, since the Weyl vector and co-Weyl vector of the Lie
algebra $G_2$ associated with the $W_{2,6}$ algebra are $\vec\rho=(\ft32,
\sqrt{\ft1{12}})$ and $\vec\rho^\vee = (5, \sqrt{3})$, that one could
express the central charges as $c=194 + 28 t^2 + 336 t^{-2}$ and $c_1 = -3t
-10 t^{-1}$.  However the solution we have found implies that this is true
only at $t=-2$, corresponding to the central charge $c=390$.  The spin--6
current $W$ is given by
\begin{eqnarray}
W&=&W_0 -\ft16 J_0^6 -\ft12 T_0 J_0^4 -\ft{4921}{114718} T_0^3
-\ft38 T_0^2J_0^2 + \ft98 T_0^2\del J_0
+ \ft{15}2 T_0\del J_0\, J_0^2
- \ft{21}2 T_0(\del J_0)^2 \nonumber\\
&&-\ft{41}4 T_0\del^2 J_0\, J_0 + \ft{21}4
T_0\del^3 J_0
+ \ft{11}2 \del J_0\, J_0^4 -\ft{315}8 (\del J_0)^2 J_0^2
+ \ft{277}8 (\del J_0)^3 + \ft74 \del T_0\, J_0^3 \nonumber\\
&&+ \ft32 \del T_0\, T_0 J_0
-\ft{57}{4} \del T_0\, \del J_0\, J_0 -\ft{190257}{229436} (\del T_0)^2
+\ft{43}4 \del T_0\, \del^2 J_0
-\ft{157}{12} \del^2 J_0\, J_0^3\label{w26real1}\\
&&+ \ft{409}4 \del^2 J_0\, \del J_0\, J_0 -\ft{1763}{48} (\del^2 J_0)^2
-\ft{108753}{114718} \del^2 T_0\, T_0 -\ft{45}{16} \del^2 T_0\, J_0^2
+\ft{135}{16}\del^2 T_0\, \del J_0\nonumber\\
&& +\ft{273}{16}\del^3 J_0\, J_0^2 -
\ft{787}{16} \del^3 J_0\, \del J_0 +\ft52 \del^3 T_0\, J_0 -
\ft{197}{16} \del^4 J_0\, J_0 -\ft{440915}{458872}\del^4 T_0 +
\ft{383}{96} \del^5 J_0\ .\nonumber
\end{eqnarray}
It may be that a linearisation of $W_{2,6}$ for arbitrary central charge is
possible if further currents are added.

     Now let us turn our attention to the study of the $W_{2,s}$ strings.
Eqn.\ (\ref{realab}) provides new realisations of the $W_{2,s}$ algebras,
for $s=1,2,3,4$ and 6.   If the current $W_0$ is zero, then the
resulting realisation is precisely the same as the free-scalar realisation,
with the distinguished scalar $\del\phi_1$ replaced by the abstract spin--1
current $J_0$.  However it was shown, for the cases of $s=3, 4$,
that the current $W_0$ does not have to be zero, and that instead it could
be realised by a parafermionic vertex operator \cite{ks1,ks2}. One can
alternatively realise $W_0$ in terms of a ghost-like field \cite{lpsx,bn}.
It was shown in Ref.\ \cite{lpsx}, by performing local canonical field
redefinitions, that for the latter realisations, with $s=1,2,3$, the
corresponding $W_{2,s}$ strings are equivalent to the bosonic string.   In
this letter, we shall construct new realisations involving ghost-like fields
for $W_{2,4}$ and $W_{2,6}$, and argue that these realisations provide
embeddings of the bosonic string in the corresponding $W$ strings.

     First let us consider the $W_{2,4}$ case.  To obtain a realisation for
the linearised $W_{1,2,4}$ algebra (\ref{genform}), we introduce a pair of
bosonic ghost-like fields $(r, s)$ with spins $(4,-3)$, and a pair of
fermionic ghost-like fields $(b_1, c_1)$ with spins $(k, 1-k)$.   The
realisation is then given by
\begin{eqnarray}
T_0&=&T_X  +4 r\del s + 3 \del r\, s - k\, b_1\del c_1
-(k-1) \del b_1\, c_1\ , \nonumber\\
W_0 &=& r\ , \qquad J_0 = -t\, rs + \sqrt{t^2 -1}\, b_1 c_1\ ,
\label{w24realgen}
\end{eqnarray}
where $(2k-1)^2=16(1 - t^{-2})$, and $T_X$ is an arbitrary energy-momentum
tensor with central charge $c_X= -13 + 30t^2 + 12 t^{-2}$.  The total
central charge of the realisation is $c=86 + 30t^2 +60 t^{-2}$.   Once
having obtained a realisation of $W_{1,2,4}$, one can obtain a realisation
for $W_{2,4}$ of the form (\ref{realab}) by a basis change \cite{ks2}. Note
that when $t^2=1$, the $(b_1, c_1)$ term is absent from $J_0$, and thus it
can be absorbed into $T_X$, giving rise to effective central charge
$c_X=30$.  In this case, the realisation takes its simplest form. The
corresponding total central charge is $c=176$.

     To quantise a $c=176$ $W_{2,4}$ string, we need to use a non-critical
BRST construction, in which we introduce a $c=-4$ Liouville sector, since
the critical central charge for the $W_{2,4}$ string is $c=172$.  The
non-critical BRST operator for $W_3=W_{2,3}$ was first obtained in Ref.\
\cite{blnw}.  Subsequently, the non-critical $W_{2,4}$ BRST operator was
constructed in Ref.\ \cite{zhao1}. The two-scalar realisation of the
$W_{2,4}$ algebra was first constructed in Ref.\ \cite{kw}.  We can instead
realise the $W_{2,4}$ algebra by two pairs of fermionic fields $(b_2, c_2)$
and $(b_3, c_3)$.  When $c=-4$, the realisation takes the particularly
simple form
\begin{eqnarray}
T_{\rm L} &=& -b_2\del c_2 -b_3\del c_3\ , \nonumber\\
W_{\rm L} &=& b_2\del c_2\, b_3\del c_3 + \ft52(T_{\rm L}^2 -
\ft3{10}\del^2 T_{\rm L})\ . \label{w24bcreal}
\end{eqnarray}
Note that this realisation does not close classically, but it does close at
the quantum level.  If one bosonises the $(b_2, c_2)$  and $(b_3, c_3)$
fields, it is equivalent to the two-scalar realisation.   With this
realisation for the $c=-4$ Liouville sector, one can write down the
non-critical $W_{2,4}$ BRST operator in the graded form \cite{zhao1}
\begin{eqnarray}
Q_0&=&\oint c\Big( T +T_{\rm L}- 4\,\beta\del \gamma
-3\del \beta\, \gamma - b\del c\Big)\ , \nonumber\\
Q_1&=&\oint \gamma\Big (195 \sqrt{\ft2{451}}W -\ft{59}{451} T^2 +
b_2\del c_2\, T -
4 b_2\del c_2\, \beta\del\gamma + T\beta\del\gamma
- \ft{298}{451} \del^2 T\label{w24brst1}\\
&& +3 b_2\del^3 c_2 + 5\del b_2\, \del^2 c_2 + 3\del^2 b_2\, \del c_2
 +  3 \beta\del^3\gamma +
2\del\beta\, \del^2\gamma\Big)\ ,\nonumber
\end{eqnarray}
where $(c, b)$ and $(\gamma, \beta)$ are the ghost fields for the
spin--2 and spin--4 currents respectively, and $T$ and $W$ generate the
$W_{2,4}$ algebra with $c=176$.  It is interesting that this non-critical
BRST operator with abstract matter currents has a simpler form than the
abstract critical $W_{2,4}$ BRST operator \cite{zhu}. Substituting the
$c=176$ realisation that we discussed above, the $Q_0$ operator has the same
form, with $T=T_X + 4r\del s + 3\del r\, s$, and the $Q_1$ operator can be
expressed as
\begin{equation}
Q_1=\oint \gamma\Big( r -\ft14 r^4s^4 + {\rm more}\Big)\ ,\label{w24brst2}
\end{equation}
modulo an overall constant factor, where the ``more'' terms are quantum
corrections to the classical terms $\gamma(r-\ft14 r^4s^4)$.   Note that the
Liouville sector enters the $Q_1$ operator only as a quantum correction.
The $Q_1$ operator is analogous to the one for the $W_3$ string, where
$Q_1=\oint \gamma (r -\ft13 r^3s^3 + {\rm quantum}\, {\rm
corrections})$ \cite{lpsx}.  It was shown that the $Q_1$ operator for
$W_3$ can be converted into a single term $\gamma r$ by a local canonical
field redefinition.  We expect that this can also be done for the $Q_1$
operator (\ref{w24brst2}) for $W_{2,4}$.  To see this, we note that the
classical terms $\gamma (r-\ft1{j} r^j s^j)$ can be converted into the
single term $\gamma r$ by the following local field redefinition
\begin{eqnarray}
r&\longleftarrow& r-\ft1j r^j s^j \ ,\nonumber\\
s&\longleftarrow& \sum_{n\ge 0} g_n r^{nj-n} s^{nj+1}\ ,\label{simtra}
\end{eqnarray}
where $g_n=n(nj+1)^{-1} g_{n-1}$ with $g_0=1$.   We expect that the operator
$Q_1$ in Eqn.\ (\ref{w24brst2}) can also be converted into the single term
by local field redefinitions at the full quantum level.   Since the
redefined $(r, s)$ and $(\beta, \gamma)$ fields then form a Kugo-Ojima
quartet, they do not contribute to the cohomology of the BRST operator.  The
cohomology of the $W_{2,4}$ BRST operator is thus equivalent to that of the
BRST operator of the bosonic string
\begin{equation}
Q=\oint c\Big( T_X -b_2\del c_2 - b_3\del c_3 - b\del c\Big )\ ,
\label{virasoro}
\end{equation}
where the central charge for $T_X$ is $c_X=30$.  Hence the new realisation
provides an embedding of the Virasoro string with $c=30-4$ in the
non-critical $W_{2,4}$ string.

       We can also construct the critical $W_{2,4}$ string using this new
realisation (\ref{w24realgen}).  When $t^2=\ft65$, the central charge of the
realisation (\ref{w24realgen}) takes the critical value $c=172$.  In this
case, $c_X=33$ and the central charge of the $(b_1, c_1)$ system is $-7$.
The critical realisation for the $W_{2,4}$ algebra can be straightforwardly
obtained by performing a basis change of the linear $W_{1,2,4}$ algebra
\cite{ks2}.  The critical BRST operator for $W_{2,4}$ can also be written in
a graded form $Q=Q_0+ Q_1$ \cite{w2s}. In terms of this realisation, the $Q_0$
operator is given by Eqn.\ (\ref{w24brst1}) with $T_{\rm L}$ omitted;
the $Q_1$ operator is given by
\begin{equation}
Q_1=\oint \gamma\Big(r-\ft14 r^4s^4 + \ft16 r^3s^3\,b_1c_1 +
{\rm quantum}\, {\rm corrections} \Big)\ .
\end{equation}
As in the case of the critical $W_3$ string discussed in Ref.\ \cite{lpsx},
we expect that this operator can also be converted into a single term
$\gamma r$.  Thus the cohomology of the critical $W_{2,4}$ BRST operator is
equivalent to that of the bosonic string with the energy-momentum tensor
$T_X + T_{c_1, b_1}$, where the central charges for $T_X$ and $T_{c_1, b_1}$
are 33 and $-7$ respectively.

      Now let us consider $W_{2,6}$ strings.  As we have shown in this
paper, with the inclusion of a spin--1 current the $W_{2,6}$ algebra can be
linearised only for central charge $c=390$.  The linearised $W_{1,2,6}$
algebra is given by Eqn.\ (\ref{genform}) with $s=6$, $c_1=11$ and $h=-1$.
A realisation can be easily obtained, given by
\begin{equation}
T_0=T_X + 6 r\del s+ 5\del r\, s\ , \qquad W_0=r\ ,
\qquad J_0 = rs\ .\label{w126real}
\end{equation}
The central charge for $T_X$ is 28.  The realisation for $W_{2,6}$ can then
be obtained by substituting Eqn.\ (\ref{w126real}) into Eqn.\
(\ref{w26real1}).   Since the critical central charge for $W_{2,6}$ is 388,
we need to use a non-critical $W_{2,6}$ BRST operator, with the Liouville
sector contributing a central charge $c=-2$.  The $W_{2,6}$ algebra becomes
degenerate at central charge $c=-2$ \cite{lpx}, in the sense that the OPE of
the spin--6 current with itself gives rise only to descendants of the
spin--6 currents.  Thus it is possible to set the spin--6 current to zero in
the non-critical BRST operator for $W^{\rm M}_{2,6}\otimes W^{\rm L}_{2,6}$
at this central charge, leading to a $c_{\rm M}=390$ non-critical $W_{2,6}$
BRST operator with a purely Virasoro Liouville sector, which is given by
\cite{lpx}
\begin{eqnarray}
Q_0&=&\oint c\Big(T + T_{\rm L }-6\beta\del \gamma - 5\del\beta\, \gamma -
b\del c\Big)\ ,\nonumber\\
Q_1&=&\oint \gamma \Big( \st{2448} \sqrt{\ft{41149461318}{13}}\, W_{\rm M} +
\st{4282}\, T_{\rm M}^3 +
\ft{1390837}{13}\, \del^2 T_{\rm M}\, T_{\rm M}\nonumber\\
&& +
\ft{1038100}{13}\, \del T_{\rm M}\, \del T_{\rm M}
+\ft{6815257}{39}\, \del ^4 T_{\rm M}
-\ft{1032462}{13}\, T_{\rm M}^2\,\beta\del\gamma\nonumber\\
&&+
\ft{4301925}{13}\, \del^2 T_{\rm M}\, \beta\del\gamma
+\ft{16634110}{13}\, \del T_{\rm M}\, \del \beta\, \del\gamma
+\ft{6653644}{13}\, T_{\rm M}\,\del^2\beta\, \del\gamma\, \gamma
\label{w26brst}\\
&&-\ft{9980466}{13}\, T_{\rm M}\,\beta\del^3\gamma\,\gamma
-\st{1433975}\,\del^4\beta\, \del\gamma + \st{2581155}\,
\del^2\beta\del^3\gamma\nonumber\\
&&-\st{1433975}\,\beta\del^5\gamma\,\gamma -
\st{1720770}\, \del\beta\, \beta\del^2\gamma\, \del\gamma\Big)\ ,\nonumber
\end{eqnarray}
where the Liouville current $T_{\rm L}$ generates the Virasoro algebra at
$c=-2$, and hence it can be realised as $T_{\rm L} = -b_1\del c_1$.
Substituting the new realisation (\ref{w26real1}) for $W_{2,6}$, with $T_0,
W_0$ and $J_0$ given by (\ref{w126real}),  we expect that the $Q_1$ operator
can be transformed into a single term $\gamma r$ by a local canonical field
redefinition. (Note that the classical terms in the $Q_1$ operators are
$\gamma(r -\ft16 r^6 s^6)$ modulo an overall constant factor.)  Thus the
cohomology of this BRST operator is equivalent to that of the Virasoro
string with energy-momentum tensor $T=T_X -b_1\del c_1$.

     To summarise, we have shown in this paper that the bosonic string can
be embedded into $W_{2,s}$ strings for $s=4, 6$, extending previous results
for $s=1,2$ and 3.  The key feature that makes the embedding possible is
that the realisation of the higher-spin current involves a term linear in a
ghost-like field.  The existence of such a linear term was implied by the
fact the $W_{2,s}$ ($s=3,4,6$) algebras can be linearised with the inclusion
of a spin--1 current.  Such a linearisation is possible for $W_{2,3}$ and
$W_{2,4}$ at all values of central charge \cite{ks1,ks2}.  In this paper, we
showed that for the case of $W_{2,6}$, the linearisation is possible only
when the central charge is 390.   We found realisations in terms of
ghost-like fields for the $W_{2,4}$ and $W_{2,6}$ algebras, and used these
new realisation to construct the corresponding $W$ strings.  We argued that
the associated BRST operators are equivalent to that of the Virasoro string.
It would be interesting to extend these results to other $W$ strings.   The
linearisation of the $W_N$ algebra has been obtained recently in Ref.\
\cite{ks3}, which may provide new realisations for the embedding of the
bosonic string.   It would also be of great interest to investigate the
nested embedding of the $W_N$ string in the  $W_{N+1}$ string.

\section*{Acknowledgements}

We would like to thank K.S.~Stelle for useful discussions.

\end{document}